\documentclass[prb,
twocolumn,
showpacs,
amsmath,
amssymb,
floatfix,
superscriptaddress
]{revtex4-1}
\usepackage{graphicx}
\usepackage{times}
\usepackage{color}
\usepackage[breaklinks]{hyperref}

\newcommand{\baz}{\begin{array}{cc}}
\newcommand{\ea}{\end{array}}
\newcommand{\be}{\begin{equation}}
\newcommand{\ee}{\end{equation}}

\begin{document}
\title{Quantum Hall effect in narrow graphene ribbons}

\author{H. Hettmansperger}
\affiliation{Institut f\"ur Theoretische Physik und Astrophysik,
  University of W\"urzburg, Am Hubland, 97074 W\"urzburg, Germany}

\author{F. Duerr}

\author{J.B.~Oostinga}

\author{C. Gould}
\affiliation{Physikalisches Institut (EP3), University of W\"urzburg,
  Am Hubland, 97074 W\"urzburg, Germany}

\author{B. Trauzettel}
\affiliation{Institut f\"ur Theoretische Physik und Astrophysik,
  University of W\"urzburg, Am Hubland, 97074 W\"urzburg, Germany}

\author{L.W.~Molenkamp}
\affiliation{Physikalisches Institut (EP3), University of W\"urzburg,
  Am Hubland, 97074 W\"urzburg, Germany}

\date{\today}

\begin{abstract}
The edge states in the integer quantum Hall effect are known to be
significantly affected by electrostatic interactions leading to the
formation of compressible and incompressible strips at the boundaries
of Hall bars. We show here, in a combined experimental and theoretical
analysis, that this does not hold for the quantum Hall effect in
narrow graphene ribbons. In our graphene Hall bar, which is only 60 nm
wide, we observe the quantum Hall effect up to Landau level index
$k=2$ and show within a zero free-parameter model that the spatial
extent of the compressible and incompressible strips is of a similar magnitude as the magnetic length. We conclude that in narrow graphene ribbons the single-particle picture is a more appropriate description of the quantum Hall effect and that electrostatic effects are of minor importance.
\end{abstract}

\pacs{73.63.--b, 81.05.ue, 73.43.--f}

\maketitle

\section{Introduction}

The discovery of graphene is intimately related to the first
measurements of the integer quantum Hall effect in this material
system.\cite{Novos2005,Zhang2005} It was observed as early as 2005
that the integer quantum Hall effect shows an anomalous sequence of
Hall plateaus in the Hall conductivity $\sigma_{xy} = 4(k+1/2)\,e^2/h$
due to the graphene-specific Berry phase that finds its origin in the
bipartiteness of the honeycomb lattice. This is a crucial difference
from quantum Hall effect measurements in more established material
systems such as two-dimensional electron gases (2DEGs) in
semiconductor heterostructures.

A second distinction between the quantum Hall effect in graphene and
that in ordinary 2DEGs is suggested by two terminal conductance studies on
narrow ribbons, \cite{PhysRevLett.107.086601,Schmidmeier} where
conductance quantization was interpreted in quantum Hall terms,
even though the ribbons are narrower than the typical separation
between edge states in a 2DEG. To confirm the validity of this
interpretation, we have fabricated a 60-nm-wide graphene Hall bar,
allowing for proper four terminal measurement of $R_{xy}$ and
confirming that quantum Hall edge states do survive even in such
narrow devices in graphene.

In order to reconcile these observations with the conventional picture
of edge-state separation, we need to recall the physics of edge states in ordinary 2DEGs. The first detailed description of
edge-state transport in the quantum Hall regime was provided by
Halperin \cite{Halperin} using a noninteracting electron
picture. While this model is useful in describing much of the relevant
physics, it was subsequently found that such a picture does not in
general properly account for the position of the edge states. A more
exact description emerges when considering the influence of
electrostatic interactions between the electrons that form the quantum
Hall state and the surrounding electrons in the substrate and the
nearby metallic gates. In ordinary 2DEGs, it is well established that
these electrostatic interactions lead to the formation of compressible
and incompressible strips.\cite{Chklovskii:1992vg,Chklovskii:1993cw} Indeed, the effect of
electrostatic interactions is of such importance that, in many cases,
the single-particle picture fails to provide even a qualitative
description of experiment. The situation is different in narrow
graphene structures, where, as our zero-free-parameter model shows,
the relevant length scale for the single-particle picture --- the
magnetic length $l_B=\sqrt{\hbar/eB}$ --- is similar to the width of
the compressible and incompressible strips.


\section{Experimental part}

\begin{figure}
	\centering
		\includegraphics{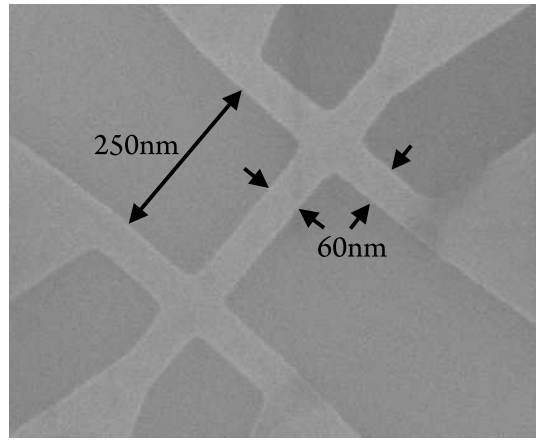}
                \caption{(Color online) SEM image of a plasma patterned six-terminal graphene
                  Hall bar with $w=60$\,nm and $l=250$\,nm.}
	\label{fig:SEM}
\end{figure}

A scanning electron micrograph (SEM) of our sample is shown in
Fig.~\ref{fig:SEM}. It is made from single-layer graphene flakes
exfoliated from an HOPG block onto a highly p-doped Si substrate
having a $285$\,nm thick SiO$_2$ cap layer. After contacting Ti$/$Au electrodes, these flakes are etched in an Ar$/$O$_2$ plasma to
obtain narrow Hall bars with six terminals. To reduce residual
contaminations, the etched structure is annealed at $200\,^\circ$C in
forming gas for about 2 h. The sample is studied at the base
temperature of a dilution refrigerator using standard low-noise ac
measurement techniques.

\begin{figure}
	\centering
		\includegraphics[width=0.45\textwidth]{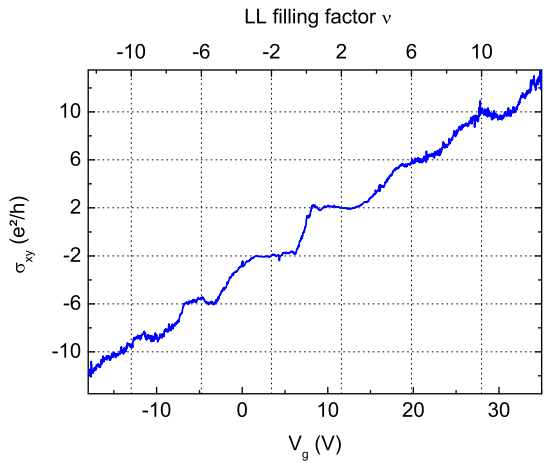}
                \caption{(Color online) Hall conductivity $\sigma_{xy}$ as a function of gate voltage $V_g$ and Landau level filling factor $\nu= h/(e^2 B) C_g(V_g-V_g^D)$ at $B=11$\,T and $30$\,mK. (Figure adapted from Ref. \onlinecite{Duerr2012}.)}
	\label{fig:sigmaxy}
\end{figure}
Preliminary characterization of the sample is done at $4.2$\,K, where,
as typical for ribbons of width below $100$\,nm, the conductance $G$
of the sample is strongly suppressed in the vicinity of the charge
neutrality point, which is at $V_g^D\approx7.5$\,V. At high electron
and hole concentrations, the field-effect mobility of the charge
carriers is $\mu=\frac{l}{2a_rC_g} \frac{dG}{dV_g} \approx
1500$\,cm$^2/$Vs, where the effective gate capacitance
$C_g\approx210$\,$\mu$F$/$m$^2$ is determined from Hall measurements
at various gate voltages $V_g$. This corresponds to a diffusion
constant of $D\simeq 0.01$\,m$^2/$s and a mean free path of $l_m
\simeq 20$\,nm. Such small values are typical for plasma-etched
graphene nanoribbon devices and are a consequence of the great amount
of edge disorder in these systems.\cite{Molitor:2011gz}

We recently observed the quantum Hall effect in very narrow graphene wires.\cite{Duerr2012} A representative example of such a measurement is shown in Fig.~\ref{fig:sigmaxy}, where we plot the measured Hall conductivity
$\sigma_{xy}$ as a function of the back-gate voltage $V_g$ and filling
factor $\nu=\pm4(k+1/2)$ at $B=11$\,T. The occurrence of conductance
plateaus at $k= 0$, $1$, and $2$ is typical for Dirac fermions and
clearly confirms that our sample is mono layer graphene.  In general,
$\sigma_{xy}$ plateaus are well developed if the energy spacing
between the corresponding Landau levels is much larger than their
broadening $\Gamma$. In graphene the former, given by
Eq.~\eqref{eq:8}, decreases with increasing filling (i.e., the index
$k$), whereas the broadening $\Gamma=\hbar v_F/l_m\simeq 30$\,meV is
constant. This explains why only the $k=0$, $1$, and $2$ plateaus are
visible up to $B=11$\,T, with the $k=0$ plateaus being most pronounced. We found this observation remarkable in view of previous experience with (Al,Ga)As--based structures. Therefore, we decided to investigate the electrostatics of the edge channels in more detail, which is presented in the next sections.


\section{Theoretical part}

From a theoretical point of view, the essence of our device is an
etched graphene ribbon separated from a metallic back gate by an
insulating SiO$_2$ layer. We note that the thickness, of the gate
insulator is a significant parameter because, for very thin thicknesses
tunneling becomes as important as electrostatic effects. Our geometry
is therefore fundamentally different from the one used in recent STM
studies of the spacial extent of edge channels in graphene resting
directly on graphite, where the thickness of the gate insulator is
basically 0 (i.e., the distance between the graphene and the graphite
substrate).\cite{Guohong}

We now examine the consequences of our geometry on the accepted models
by Chklovskii, Shklovskii, and Glazman (CSG) \cite{Chklovskii:1992vg}
as well as Chklovskii, Matveev, and Shklovskii (CMS).\cite{Chklovskii:1993cw} In the electrostatic theories of CSG and
CMS, a 2DEG is laterally confined by electric fields originating from
gate electrodes which are in plane with the 2DEG. In the absence of
magnetic fields, the electrostatic solution of such a two-dimensional
problem was provided by Larkin and Shikin.\cite{Larkin:1990wr} They
find a charge density distribution which reduces from its bulk value
to 0 at the boundary of the electron gas. Although the confinement
of mobile electrons in these models is generated by gates, the authors
of Refs.~\onlinecite{Chklovskii:1993cw,Chklovskii:1992vg} argue that their
results also apply to etched structures of conventional
heterostructures. This is justified by surface states which lead to
Fermi level pinning and negative charges accumulate at the edges.
Inner electrons are repelled, yielding the same depletion effect as
split gates. However, as shown by Silvestrov and Efetov (SE),
\cite{Silvestrov:2008bk} the situation is substantially different in
the case of a graphene nanoribbon. Since the edges of this
two-dimensional 
material end abruptly, they act as hard- instead of
soft-wall boundaries. Assuming a constant electrostatic potential in a
ribbon positioned in the $xz$ plane, such that in the transverse
$x$ direction no electric force acts on its excess electrons, SE
derived an expression for the charge density. If the ribbon width
$2a_r$ is smaller than the gate dielectric thickness $b$, the electron density
across the ribbon is, to a good approximation, given by
\cite{Silvestrov:2008bk}
\begin{equation}
  \label{eq:1}
  \rho(x,y) = \frac{\sigma}{\pi}\frac{\delta(y)}{\sqrt{a_r^2-x^2}}\left\{1-C_2\left(\frac{a_r}{b}\right)^2\left(\frac{2x^2}{a_r^2}-1\right)\right\}\,,
\end{equation}
where $ C_2 \equiv \sum_{n=1}^{\infty}
\frac{\epsilon}{1-\epsilon}\,\left(\frac{1-\epsilon}{1+\epsilon}\right)^n\frac{1}{4n^2}$ is
a numerical constant \cite{Note1}
 that, for a dielectric constant $\epsilon \simeq 3.9$ as in the case of ${\rm SiO}_2$, has a value of 0.175.  The charge per unit length
$\sigma$ is linearly related
to the back-gate potential $V_g$,\cite{Silvestrov:2008bk}
\begin{equation}
  \label{eq:3}
  V_g =\frac{4\sigma}{\epsilon+1}\left\{
    \ln{\frac{4b}{a_r}} + C_0+\left(\frac{a_r}{b}\right)^2\,C_2\right\}\,,
\end{equation}
with
 $ C_0 \equiv
 \sum_{n=2}^{\infty}\frac{2\epsilon}{1-\epsilon}\left(\frac{1-\epsilon}{1+\epsilon}\right)^n\ln
 n$.

 The inverse-square-root edge singularity in Eq.~\eqref{eq:1} is in stark contrast to the soft density profiles
 ($\rho\sim\sqrt{x}$) predicted at the edges of conventional
 heterostructures.\cite{Chklovskii:1993cw,Chklovskii:1992vg,Larkin:1990wr} In the
 following, we analyze the impact of this diverging density
 distribution on the formation of compressible and incompressible
 strips in graphene.

\begin{figure*}
  \centering
  \includegraphics[width=\textwidth]
{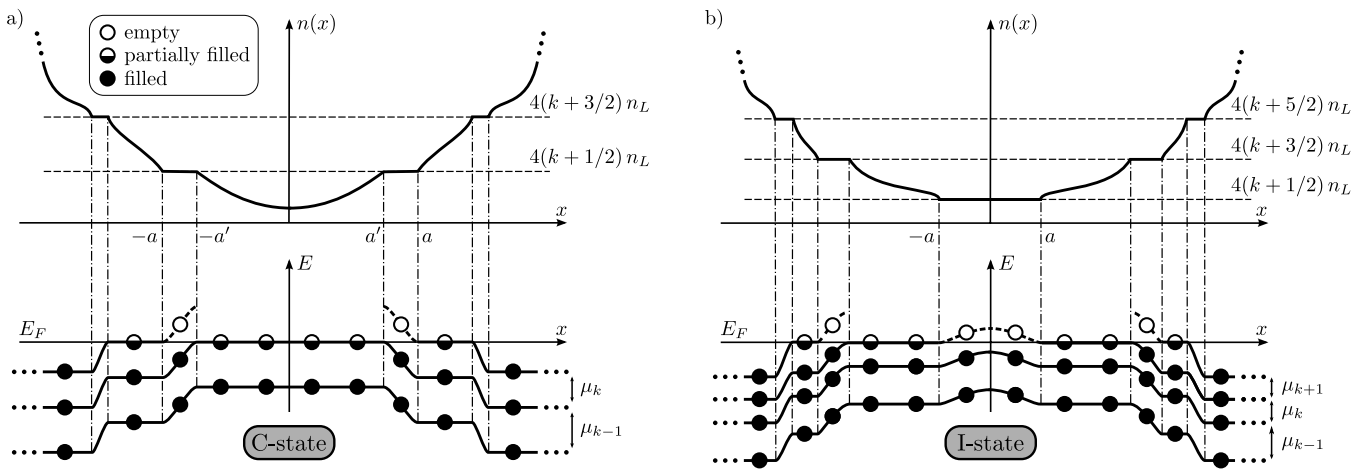}
\caption{(Color online) Schematic of the compressible and incompressible strips in graphene, showing the
  electron density $\propto n(x)$ and the Landau levels with their
  local fillings. In (a) the ribbon is in a C state with a
  compressible strip at its center whereas in (b) it is in an I state with
  an incompressible strip in the middle. The regions near the edges of the ribbon where the Landau levels
  increase very sharply due to hard-wall confinement are not shown.
  In these regions of size $\sim l_B$ away from the edges, the
  electrostatic analysis is no longer applicable.\cite{Note2}
  }
  \label{fig:C-I-state}
\end{figure*}
The formulas above are valid for any metallic strip with a hard-wall
boundary. The graphene-specific physics comes into play with quantum
effects. Evidently, the Pauli principle prevents the excess electrons
from all being in the same state. When the Fermi wavelength varies
slowly on the considered length scale (here the width of the ribbon),
one can divide the system into many elements of sizes where
equilibrium thermodynamics is still applicable. By using this
Thomas-Fermi approximation, SE defined a local Fermi momentum for each
element and derived a relation between the inhomogeneous charge
density given by Eq.~\eqref{eq:1} and the potential felt by an
individual electron. At zero temperature and in the limit of rather
narrow ribbons ($a_r \ll b$), the curly bracket in Eq.~\eqref{eq:1}
reduces to $1$, and the potential has the form
\begin{equation}
  \label{eq:5}
  U(x) =
  -\hbar\,v_F\,\sqrt{\sigma/e}\left(\frac{1}{a_r^2-x^2}\right)^{1/4}\,,
\end{equation}
with $v_F$ the Fermi velocity in graphene. The derived expression is
valid at distances $\delta x \gtrsim
a_r^{1/3}\left(\frac{e}{\sigma}\right)^{2/3}$ from the strip edges, where the Thomas-Fermi approximation is applicable.  This
potential should be viewed as a quantum correction to the electrostatic
potential and describes locally the energy difference between the
constant Fermi energy and the Dirac point.

Knowing the electron distribution and the potential in a graphene
ribbon, we can now adapt the CMS formalism to graphene. Making use of
the semiclassical electrostatic picture described above, we expect,
for a finite magnetic field, the
scenario sketched in Fig.~\ref{fig:C-I-state}.  As in the case of the
2DEG in Ref.~\onlinecite{Chklovskii:1993cw}, the electron density $\sim
n(x)$ in our graphene ribbon is divided into alternating compressible
and incompressible strips. The latter are characterized by filled
Landau levels and describe local unscreened regions in which electrons
have no possibility to reorder. This is due to the large energy gap
they must overcome in order to find unoccupied states in one of the
nearby empty levels. On the other hand, compressible strips are
characterized by partially filled Landau levels residing at the Fermi
level.  The many unoccupied states allow for the electrons to
rearrange and minimize their energy such that the electric field is
locally screened. In general, two situations are possible. In
Fig.~\ref{fig:C-I-state}(a), we show a C state
that describes a central compressible strip in the $k$-th Landau
level which is partially filled near $x=0$. Upon decreasing the magnetic
field strength or increasing the density of excess charges, the $k$-th
Landau level gets filled and the ribbon is driven into an I state with
an incompressible strip at the center [cf.~Fig.~\ref{fig:C-I-state}(b)].
\begin{table}[b]
\caption{\label{tab:1} Experimental parameters used in the modeling.}
\begin{center}
\begin{tabular}{l|r}
  \hline
 Quantity & Value  \\
  \hline
Magnetic field $B$ & $11\,\mathrm{T}$ \\
Insulator dielectric constant $\epsilon$ & $3.9$, $\mathrm{SiO}_2$ \\
Thickness of insulator $b$ & $0.3\,\mu\mathrm{m}$ \\
Half-width of graphene ribbon $a_r$ & $30\,\mathrm{nm}$\\
Back-gate potential $V_g$ & $-20$ to $35$ V\\
   \hline
\end{tabular}
\end{center}
\end{table}

\section{Results}

The basic formalism used in Ref.~\onlinecite{Chklovskii:1993cw} to determine the widths and
positions of the compressible and incompressible strips in a gate
confined 2DEG is also valid in
graphene nanoribbons. Imposing the same requirements on the
electrostatic boundary conditions as within the framework of the CMS
model, and solving for our geometry leads to the following set of equations:
\begin{align}
  \label{eq:1a}
  \left[\nu(0)-4(k+1/2)\right]n_L+n''(0)\frac{a^2+a'^2}{4} &= 0\,,
\end{align}
\begin{multline}
  -\frac{2\,\pi\, e\, n''(0)\,
    a}{3(\epsilon+1)}\left\{(a^2+a'^2)E\left[\sqrt{1-(a'/a)^2}\right]-\right.\\\left.-2(a')^2K\left[\sqrt{1-(a'/a)^2}\right]\right\}
  = -\mu_k\,,\label{eq:1b}
\end{multline}
\begin{align}
\nu(0)-\nu_B(0) &= -\frac{n''(0)(a-a')^2}{4n_L}\nonumber\\ &= \frac{(a-a')^2}{a^2+a'^2}\left[\nu(0)-4(k+1/2)\right]\,.\label{eq:1c}
\end{align}
Here, $E$ and $K$ are elliptic integrals, and the quantities $a$ and
$a'$ determine the positions and widths of the compressible and
incompressible strips next to $x=0$. The
energy difference,
\begin{equation}
\mu_k\equiv
 \frac{\hbar v_F}{l_B}\left(\sqrt{2(k+1)}-\sqrt{2k}\right)\,,\label{eq:8}
\end{equation}
is that of adjacent Landau levels (see Fig.~\ref{fig:C-I-state}) and
depends on the index $k$, the Fermi velocity $v_F$, and the magnetic
length $l_B=\sqrt{\hbar/eB}$. The occupation number of these levels is
defined by $\nu_B(0) \equiv n_B(x=0)/n_L$, with $n_B(0)$ the electron
concentration at magnetic field strength $B$ in the center of the
ribbon, and $n_L \equiv 1/2\pi l_B^2$ the Landau level degeneracy.
Furthermore, we define $\nu(0) \equiv n(x=0)/n_L$, with $n(x)$ the
electron concentration in the absence of a magnetic field, which can be
extracted from $\rho(x,y) = e\,n(x)\, \delta(y)$ via Eq.~\eqref{eq:1}.

\begin{figure}[t]
  \centering
  \includegraphics[width=0.43\textwidth]{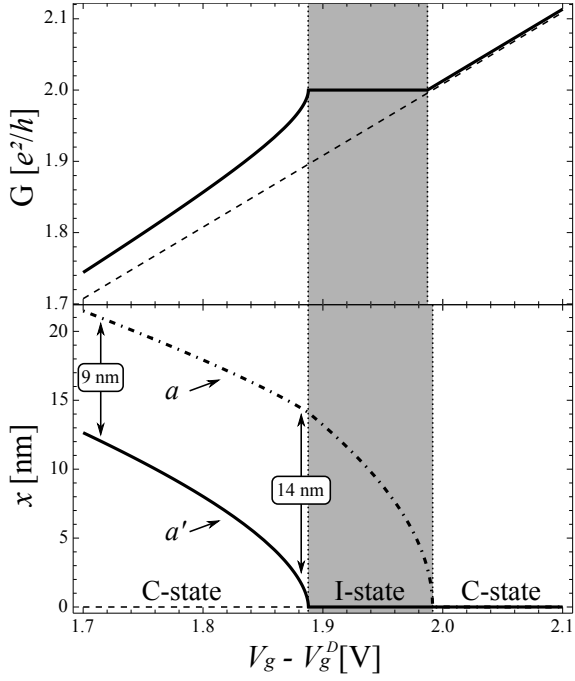}
  \caption{(Color online) Top: Two-terminal conductance plateau
    corresponding to Landau level index $k=0$, plotted over the back-gate
    voltage $V_g$. The charge neutrality point is at $V_g^D$, and the
    dashed line indicates $G_0$. Bottom: The position and width of the
    compressible and incompressible strips are plotted, which, in
    Fig.~\ref{fig:C-I-state}, correspond to the occupation level
    $4(k+1/2)\,n_L$ with $k=0$. The dashed-dotted line shows $a$, the
    outer edge position; the solid line shows $a'$, the inner edge
    position of the incompressible strips. In the upper graph the
    suppressed contribution of distant incompressible strips to the conductance
    at the I-C transition is incorporated, whereas in the lower
    graph it is neglected, resulting in a slight mismatch in width of the
    shaded areas.
    }
  \label{fig:plateau}
\end{figure}
Taken together, Eqs.~\eqref{eq:1a}, \eqref{eq:1b}, and \eqref{eq:1c}
represent a complete system from which the quantities $a$, $a'$, and
$\nu_B(0)$ can be derived. The latter is linearly related to the two-terminal conductance \cite{Chklovskii:1993cw} via
\begin{equation}
  \label{eq:7}
  G = \frac{e^2}{h}\nu_B(0)\,.
\end{equation}
In order to analyze the results provided by this model, we use the experimental parameters listed in
Table~\ref{tab:1}, and solve for $a$, $a'$, and $\nu_B$.
Since Eqs.~\eqref{eq:1a}, \eqref{eq:1b}, and \eqref{eq:1c} are
derived under the assumption that $n(x) \simeq n(0)+\frac{1}{2}n''(0)\,x^2$, we are
limited to a parameter range where this
approximation is valid. This is ensured when $a$ and $a'$ are
small compared to $a_r$ which is the case at the transition from a C
to an I state caused by increasing the doping or decreasing the
magnetic field. In the upper graph in Fig.~\ref{fig:plateau}, we show the
conductance plateau which corresponds to the Landau level with index
$k=0$. From the electrostatic point of view, incompressible strips
next to $x=0$, characterized by local regions with an additional
charge accumulation, are described by dipolar strips.\cite{Chklovskii:1992vg} Although the calculated charge density
distribution, Eq.~\eqref{eq:1}, is not altered on a rough scale at any
reasonable magnetic field, the formation of such strips leads to a
small enhancement in charge concentration at $x=0$. Thus, it is reasonable that the conductance,
given by Eq.~\eqref{eq:7}, lies above the dashed line, which indicates
$G_0\equiv\frac{e^2}{h}\nu(0)$. From the lower graph in Fig.~\ref{fig:plateau}, we can read
off the positions and widths of the compressible and incompressible
strips which correspond to the occupation level $4(k+1/2)\,n_L$ with
$k=0$.  In the C state, before the formation of the conductance
plateau, the incompressible strips next to the central compressible
one have a width of $\sim 9\text{ nm ... }14\text{ nm}$. This, however,
contradicts a crucial assumption of the electrostatic model,\cite{Chklovskii:1993cw}
\begin{equation}
  \label{eq:9}
  l_B\sqrt{k} \ll (a-a')\,.
\end{equation}
For the quasi classical electrostatic treatment above to be
justified, the size of the electron wave function determined by the
magnetic length should be much smaller than the length scales provided
by the widths of the compressible and incompressible strips. Since
$l_{B=11T} \sim 8$ nm, our experiment is in the crossover regime
to the conventional single-particle picture of ballistic transport.\cite{1990PhRvB..41.7906B} The latter is mandatory at high quantum
numbers $k$ where condition \eqref{eq:9} is strongly violated. For the
first several plateaus, however, the electrostatic model which results
in Eqs.~\eqref{eq:1a}, \eqref{eq:1b}, and \eqref{eq:1c} is still relevant, as Eq.~(\ref{eq:9}) is approximately valid. Also note that the widths of incompressible strips
located at $x_k$
scale as $\sim1/\sqrt{n'(x_k)}$.\cite{Chklovskii:1992vg} Thus, for narrow
ribbons with large $n'(x_k)$ and therefore small $a-a'$, condition
\eqref{eq:9} is violated much earlier as for wide ribbons with
expanded incompressible strips. Furthermore, since $n'(x_k)$ increases
as we approach the edges, the spatial extent of the edge states within
the electrostatic picture is much narrower in our graphene ribbon
compared to $l_B$. Hence, we show that the CSM picture of
transport in the quantum Hall regime breaks down here and the simpler single-particle
picture not only suffices but, indeed, is essential to correctly describe the physics.

\section{Comparison of theory and experiment}

Comparing the data on the $k=0$ plateau, corresponding to filling
factor $\nu=2$ in Fig.~\ref{fig:sigmaxy}, with the theoretical result
given in Fig.~\ref{fig:plateau}, we find discrepancies in width and
location, both measured in units of the back-gate voltage. Our
electrostatic model leads to very
narrow plateaus occurring at gate voltages which are too small by a
factor of $\sim3$ to $4$. This disagreement is consistent with theory overestimating the gate capacitance, which is a commonly observed discrepancy in graphene nanoribbons.\cite{Lian2010}
The narrowness of the theoretically derived plateaus compared to their experimental widths
can be explained by disorder, which is certainly present in the device, but
neglected in our model. As realized in
Refs.~\onlinecite{Chklovskii:1993cw,Chklovskii:1992vg}, disorder leads to
localization in a compressible liquid a low small enough density, such
that at the I to C state transition there is a range of voltage
where Landau levels which locally (in the vicinity of $x=0$) start to
fill up cannot contribute to the conductance.

\section{Conclusions}

We have performed four terminal transport measurements on
a 60-nm-wide Hall bar of graphene. Our results validate the interpretation that the
Hall effect is responsible for the quantized conductance in narrow
ribbons. A quantitative study of the electrostatics of graphene
nanoribbons, and its implication for the spatial extent of edge states in such devices,
confirms that, indeed, the formation of compressible and incompressible
strips which dominate the physics of traditional 2DEG devices is
of minor importance here. Instead, the single-particle picture is more
appropriate.

\acknowledgements

B.T. would like to thank Misha
Fogler for interesting discussions at the KITP in Santa Barbara. This work was supported by the DFG in the framework of the strategic Japanese-German cooperative program and the European Science Foundation (ESF) under the EUROCORES Program EuroGRAPHENE.



\providecommand{\href}[2]{#2}

\end{document}